\documentstyle[11pt,paspconf]{article}

\begin{document}

\title{NEAR-INFRARED PHOTOMETRY OF BLAZARS}

\author{C. Chapuis\altaffilmark{1}, S. Corbel and P. Durouchoux}
\affil{Service d'Astrophysique DAPNIA, CEA Saclay F-91191 Gif sur Yvette cedex}

\author{W. Mahoney and T. N. Gautier}
\affil{Jet Propulsion Laboratory 169-327, 4800 Oak Grove Dr., Pasadena, CA 91109}

\altaffiltext{1}{D\'epartement de physique, Universit\'e de Versailles,
F-78035 Versailles cedex}

\begin{abstract}

Two days of observations were conducted
at the Palomar Observatory during the nights of 25 and 26 February 1997 with the
Hale telescope, in order to search for rapid variability in the near-infrared
(NIR) bands J, H, K$_{s}$ for a selection of eight blazars.
With the possible exception of PKS 1156+295, no intraday or day-to-day
variability was observed during these two nights.  However, for these
eight blazars, we have measured the NIR spectral indices.

\end{abstract}


\section{Introduction}

The discovery that blazars (i.e., optically violently variable 
quasars and BL Lac objects) and flat radio-spectrum quasars emit 
most of their power in high-energy gamma rays (Fichtel, et al. 1994) 
probably represents one of the most surprising results from the Compton
Gamma-Ray Observatory (CGRO). Their luminosity above 100 MeV in some cases
 exceeds 10$^{48}$ ergs s$^{-1}$ (assuming isotropic emission) and can
 be larger (by a factor of 10-100) than the luminosity in the rest of the 
electromagnetic spectrum. Blazars 
have smooth, rapidly variable, polarized continuum emission from radio 
through UV/X-ray wavelengths. All have compact flat-spectrum radio cores, 
and many exhibit superluminal motions.

A strong correlation between gamma-ray and near-infrared luminosities was 
recently discovered for a sample of blazars and it was 
suggested that this relation might be a common property of these objects 
(Xie et al. 1997). For that reason, they conclude that hot dust is likely 
to be the main source of the soft photons (near-infrared), which are 
continuously injected within the knot and then produce $\gamma$-ray flares 
by inverse Compton scattering on relativistic electrons.

The aim of the following observations was to search for an intraday or a
 day-to-day NIR variability in a sample of blazars. We expose the results
of two consecutive days observation run.

\section{Observations}

We observed eight blazars
with the 5-meter Hale telescope on Mt. Palomar during the nights of 25 and 26 
February 1997, using the Cassegrain Infrared Camera, a $256 \times 256$-pixel
InSb array with the J (1.25 $\mu$m), H (1.65 $\mu$m) and K$_{s}$ 
(2.15$\mu$m) filters and a field-of-view of 32 arcsec. 

The reduction of data was done under IRAF and included subtraction of 
the dark noise, flat field corrections, and combination of images to
remove bad pixels, cosmic rays, and the sky. Then aperture photometry for each
object was performed using nearby faint standards for calibration.
The apparent magnitudes at Earth are summarized  in Table 1.

With the possible exception of PKS 1156+295, no intraday or day-to-day
variability was observed during these two nights.
 Due to the steadyness of the sources, it was possible to fit
the data to a powerlaw (f(E) = C E$^{-\alpha}$) by a $\chi^{2}$ minimization.
We give the spectral index for each source in Table 1.
Further discussion about the results can be found in Chapuis et al. (1998a,b).

\begin{table}
\begin{center}\scriptsize
\begin{tabular}{|c|ccc|c|}
\hline
 & & & & \\
IAU name & & ${F}_{\nu}$ (mag) &  & $\alpha$ \\
 & J & H & K$_s$ &  \\
\hline
\hline
0446+112 &   17.54(10) &  16.62(6) &  15.66(5)  & 1.6 \\
\hline
0528+112 &   16.51(5)  &  15.67(5) &  14.77(5)  & 1.4 \\
\hline
0716+714 &  11.97(5)  &  11.19(5) &  10.44(5)  & 1.0 \\
\hline
0804+499 &   16.66(5)  &  15.81(5) &  15.27(5)  & 1.0 \\
\hline
0836+710 &   15.63(5)  &  15.05(5) &  14.33(7)  & 0.6 \\
\hline
1156+295 &   14.93(5)  &  14.18(5) &  13.43(5)  & 0.9 \\
         &   14.64(5)  &  13.88(5) &  13.22(5)  & 0.8 \\
\hline
1253-055 (3C279)&14.02(3)& 13.05(5) &  12.14(5)  & 1.6 \\
\hline
1406-076 &   16.96(5)  &  16.02(5) &  15.26(5)  & 1.3 \\
\hline
\end{tabular}
\end{center}
\caption{The meaning of each column for each target is as follows:
 IAU name, apparent magnitudes at Earth for each object in three bands,
  spectral index $\alpha$ in NIR range}
\end{table}

\acknowledgements Observations at the Palomar Observatory were made as part of a
continuing collaborative agreement between Palomar Observatory and the
Jet Propulsion Laboratory.  The research described in this paper was carried out
in part by the Jet Propulsion Laboratory, California Institute of Technology,
under contract to the National Aeronautics and Space Administration.

\end{document}